\documentclass{icrc2009}

\usepackage{graphicx}   % for including figures
\usepackage{caption}    % for captions
\usepackage[font=footnotesize]{subfig} % subfig.sty for a double column floating figure using two subfigures
\usepackage{fixltx2e}
\usepackage{url}

\newcommand{\shorttitle}[1]%
{\markboth{Proceedings of the 31\MakeLowercase{$^{st}$} ICRC, {\L}\'{o}d\'{z} 2009}{#1} }
\newcommand{\etal}{\MakeLowercase{\textit{et al. }}} % "et al."

\usepackage{epstopdf}

\begin{document}
\title{The strong flaring activity of M\,87 in early 2008 as observed by the MAGIC telescope}

\author{\IEEEauthorblockN{
D. Tescaro\IEEEauthorrefmark{1}, 
D. Mazin\IEEEauthorrefmark{1},
R. M. Wagner\IEEEauthorrefmark{2}, 
K. Berger\IEEEauthorrefmark{3}
N. Galante\IEEEauthorrefmark{2}\\
on behalf of the MAGIC collaboration}
\\
\IEEEauthorblockA{\IEEEauthorrefmark{1}IFAE, Edifici Cn., Campus UAB, E-08193 Bellaterra, Spain}
\IEEEauthorblockA{\IEEEauthorrefmark{2}Max-Planck-Institut f\"ur Physik, D-80805 M\"unchen, Germany}
\IEEEauthorblockA{\IEEEauthorrefmark{3}University of \L\'od\'z, PL-90236 Lodz, Poland}
}

% please write the preseter's name and short title (3-4 words maximum)
%    which will appear at the header of the even pages.
\shorttitle{D. Tescaro \etal The MAGIC detection of M\,87 flaring activity in early 2008}
\maketitle

\begin{abstract}
M\,87 is the first known radio galaxy to emit very high energy (VHE) gamma-rays. During a monitoring program of M\,87, a rapid flare in VHE gamma-rays was detected by the MAGIC telescope in early 2008 \cite{M87}. 
The flux was found to be variable on a timescale as short as 1 day, reaching 15\% of the Crab Nebula flux above 350~GeV. 
In contrast, the flux at lower energies (150 GeV to 350 GeV) is compatible with being constant. 
We present light curves and energy spectra, and argue that the observed day-scale flux variability favours the M\,87 core as source of the gamma-ray emission rather than the bright know HST-1 in the jet of M\,87.
\end{abstract}

 
\section{Introduction}
M\,87 is a giant elliptical radio galaxy (RG) of Fanaroff Riley I type (FR I; \cite{Fanaroff_Riley}) in the Virgo Cluster at a distance of 16~Mpc \cite{Macri}. 
It is powered by a supermassive black hole (BH) of $(3.2\pm0.9)\times10^{9} \mathrm{M}_\odot$ \cite{Macchetto}. 
The M\,87 jet was the first-ever observed \cite{Curtis}, and due to the proximity of M\,87, its morphological substructures can be resolved and a unique view of its innermost regions is possible. 
The jet, originating from the RG core, extends to 20'' (\cite{Marshall}, equivalent to a 2 kpc projected linear distance).
Several compact regions (ÒknotsÓ) along its axis are resolved in the radio, optical, and X-ray regimes. 
These knots have similar morphologies in all wave bands, although the X-ray knots appear to be tens of pc closer to the core than the optical and radio knots \cite{Wilson}. 
The variable brightness of the knots may be due to several shock fronts in the jet, being responsible
for particle acceleration and nonthermal emission. 
Superluminal motion of the knots has been observed in the optical \cite{Biretta} and radio \cite{Forman} wave bands, constraining the viewing angle of the jet to $<43^\circ\pm4^\circ$.
The innermost resolved bright knot HST-1 is located at 0.82'' (64~pc) from the core and is the most prominent feature of the jet. 
HST-1 has shown many flares exceeding the luminosity of the M\,87 core emission. 
Its X-ray brightness has increased by a factor 150 from 2000 to 2005 \cite{Harris}. 
A correlation between radio, optical, and X-ray luminosity points to a common origin of the emission. 
The measured superluminal motion in HST-1 is higher than in other knots, suggesting a viewing angle of $<19^\circ$ for this part of the jet. 
The core itself is variable, too, and also shows a correlation between the emission levels from radio frequencies through X-rays \cite{Perlman}.
M\,87 was not detected by EGRET. 
The first hint of very high energy (VHE; $>$~250 GeV) $\gamma$-ray emission was reported by Aharonian et al. (\cite{aha1}, 2003), and later confirmed by Aharonian et al. (\cite{aha2}, 2006) and Acciari et al. (\cite{acc1}, 2008). 
The emission is variable on a timescale of years. 
The reported $\simeq$2 day variability \cite{aha2} narrows down the size of the emission region to be on the order of the light-crossing time of the central BH. 
With its expected low accretion rate, the M\,87 core radiation is not strong enough to attenuate significantly TeV $\gamma$-rays even at 5 Schwarzschild radii ($R_S$) away from the BH \cite{NeronovAharonian}. 
All this implies a production region in the immediate vicinity of the M\,87 core.
During later observations, no significant flux variation was found \cite{acc1}. 
An X-ray--VHE $\gamma$-ray correlation is expected in most emission models, but was not unambiguously found so far.
Whereas Aharonian et al. \cite{aha2} claim a hint of a correlation between the soft (0.3--10 keV) X-rays at HST-1 and the VHE $\gamma$-rays, Acciari et al. (\cite{acc1} 2008) find a year-by-year correlation between the 2--10 keV X-ray flux of the M\,87 core and the VHE $\gamma$-ray emission instead, but do not observe a correlation between the two energy bands on shorter timescales.
The radio to X-ray emission of the jet is due to nonthermal synchrotron radiation of relativistic electrons in the jet. 
The observed knots and flares in M\,87 point to a complicated morphology with several shock fronts producing these electrons. 
While the majority of the currently known extragalactic VHE $\gamma$-ray emitters are blazars, M\,87 is assumed to be a blazar not aligned to our line of sight \cite{Tsvetanov}.
If the observed VHE emission from M\,87 is associated with the innermost part of its jet, then blazar emission models may hold. 
Recently the H.E.S.S. collaboration reported the discovery of VHE radiation also from Centaurus\,A \cite{cenA}, another FR~I radio galaxy similar to M\,87 under various aspects. 
In blazars without prominent disk or broad-line features, the VHE emission is explained by inverse Compton processes involving the synchrotron photons and their parent electron population (synchrotron self-Compton models; e.g., \cite{Maraschi}). 
Alternatively, in hadronic models, interactions of a highly relativistic jet outflow with ambient matter \cite{Dar}\cite{Beall}, proton-induced cascades \cite{Mannheim}, or synchrotron proton radiation \cite{Muecke}\cite{aha3} may produce VHE $\gamma$-rays. 
In such a scenario, M\,87 might also account for parts of the observed ultrahigh-energy cosmic rays \cite{Protheroe}. 
It should be noted that for M\,87 the location of the VHE emission is still uncertain.
Specific emission models for high-energy processes close to the core \cite{Georganopoulos}\cite{Ghisellini}\cite{Lenain}\cite{Tavecchio}, in the large-scale jet \cite{Stawarz1}\cite{Honda}, or in the vicinity of a BH \cite{NeronovAharonian}\cite{Rieger} have been developed.
The MAGIC collaboration performed monitoring observations of M\,87 starting from 2008 January, sharing the task with the VERITAS and H.E.S.S. experiments. 
Here we report on MAGIC results from a subset of these data, revealing a variability timescale of M\,87 of 1 day. 
The energy spectrum and light curves are discussed.

\section{Observations and data analysis} 
\begin{figure}[!t]
  \centering
  \includegraphics[width=2.5in]{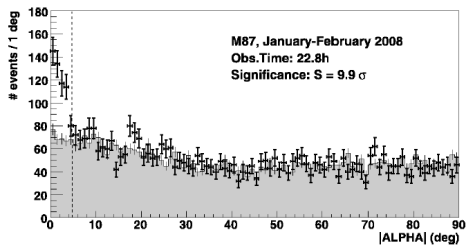}
  \caption{$|ALPHA|$ distribution for the overall data sample.
The background (gray histogram) is estimated using three OFF
regions arranged symmetrically to the ON-source region with
respect to the camera center. A $\gamma$-ray excess with a
significance of 9.9 standard deviations is obtained.}
  \label{fig1}
 \end{figure}
The MAGIC telescope is located on the Canary Island of La Palma (2200 m above sea level, $28^\circ$45'N, $17^\circ$54'W).
MAGIC is a stand-alone imaging air Cerenkov telescope (IACT) with a 17 m diameter tessellated reflector dish. 
MAGIC has a low-energy threshold of 50--60 GeV (trigger threshold at small zenith angles). 
The accessible energy range extends up to tens of TeV with a typical energy resolution of 20\%--30\%, depending on the zenith angle and energy \cite{Albert2008a}.
The data set comprises observations from 2008 January 30 to 2008 February 11. 
These were performed in the wobble mode \cite{Daum} for 26.7 hr. 
The zenith angle of the observations ranges from $16^\circ$ to $35^\circ$. 
After removing runs with unusually low trigger rates, mostly caused by bad weather conditions, the effective observing time amounts to 22.8 hr.
The data were analyzed using the MAGIC standard calibration and analysis \cite{Albert2008a}. 
The analysis is based on image parameters \cite{Hillas}\cite{newtiming} and the random forest (RF; \cite{Albert2008b}) method, which are used to define the so-called hadronness of each event. 
The cut in hadronness for $\gamma$/hadron separation was optimized on a contemporaneous data set of the Crab Nebula. 
After this cut the distribution of the angle ALPHA, which is the angle between the main image axis and the line between center of gravity of the image and the source position in the camera, is used to determine the signal in
the ON-source region. 
Three background (OFF) sky regions are chosen symmetrically to the ON-source region with respect to
the camera center. 
The final cut $|ALPHA|$$<5^\circ$ (Fig. 1) was also optimized on the Crab Nebula data to determine the number
of excess events and the significance of the signal.
The energies of the $\gamma$-ray candidates were also estimated using the RF method.
To derive a differential energy spectrum, we applied looser cuts than those in Figure \ref{fig1} to retain a higher
number of $\gamma$-ray candidates and to lower the effective analysis threshold down to 150 GeV. 
Looser cuts also reduce systematic uncertainties between data and Monte-Carlo events, which is
important for the estimation of the effective collection areas.
The derived spectrum was unfolded to correct for the effects of the limited energy resolution of the detector \cite{Albert2007}. 

% \begin{figure}[!t]
%  \centering
%  \includegraphics[width=2.5in]{figu2.eps}
%  \caption{$|ALPHA|$ distribution for the overall data sample.
%The background (gray histogram) is estimated using three OFF regions arranged symmetrically to the ON-source region with respect to the camera center. A $\gamma$-ray excess with a significance of 9.9 standard deviations is obtained.}
%  \label{fig2}
% \end{figure}
 %
  \begin{figure}[!t]
  \centering
  \includegraphics[width=2.5in]{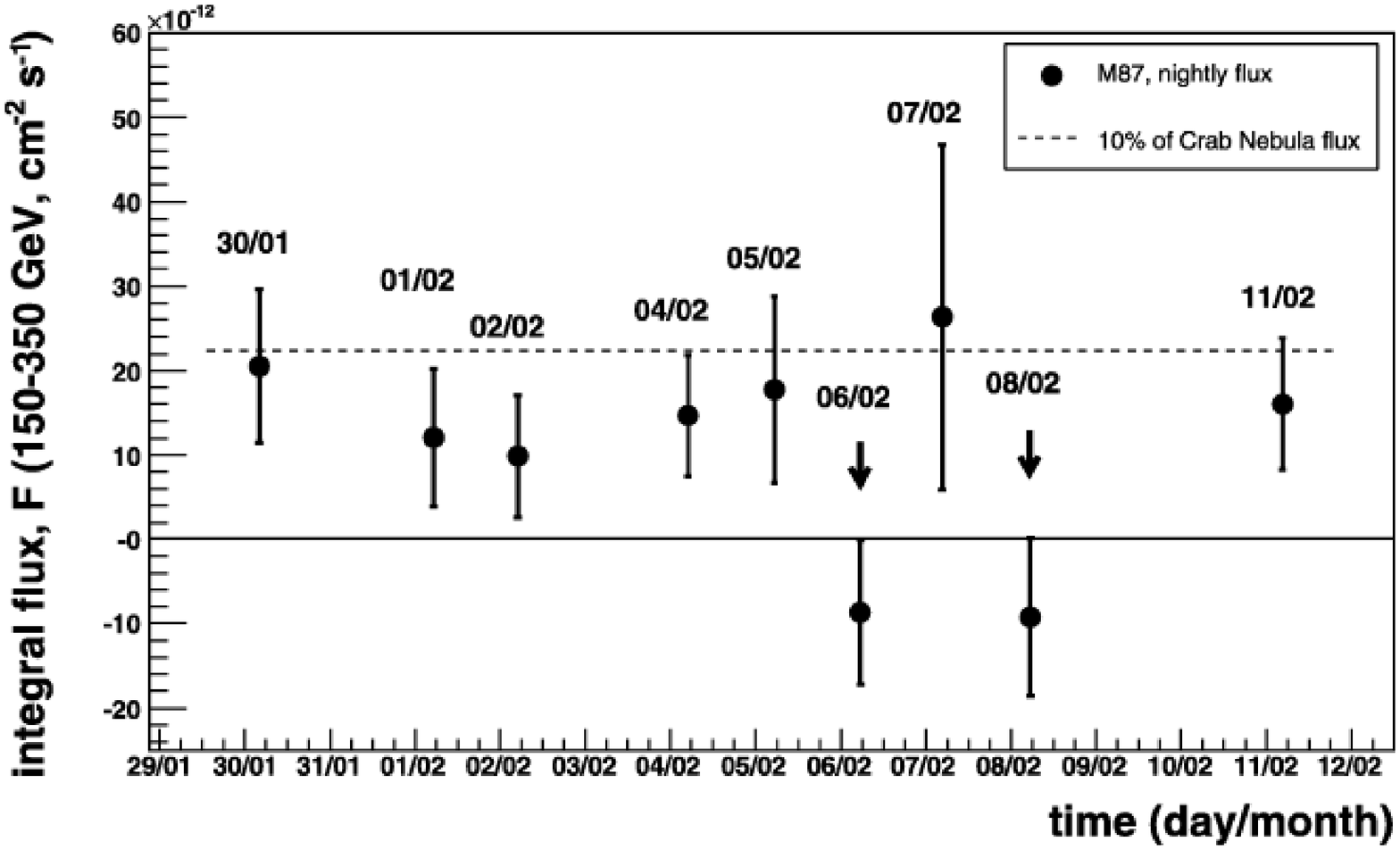}
  \includegraphics[width=2.5in]{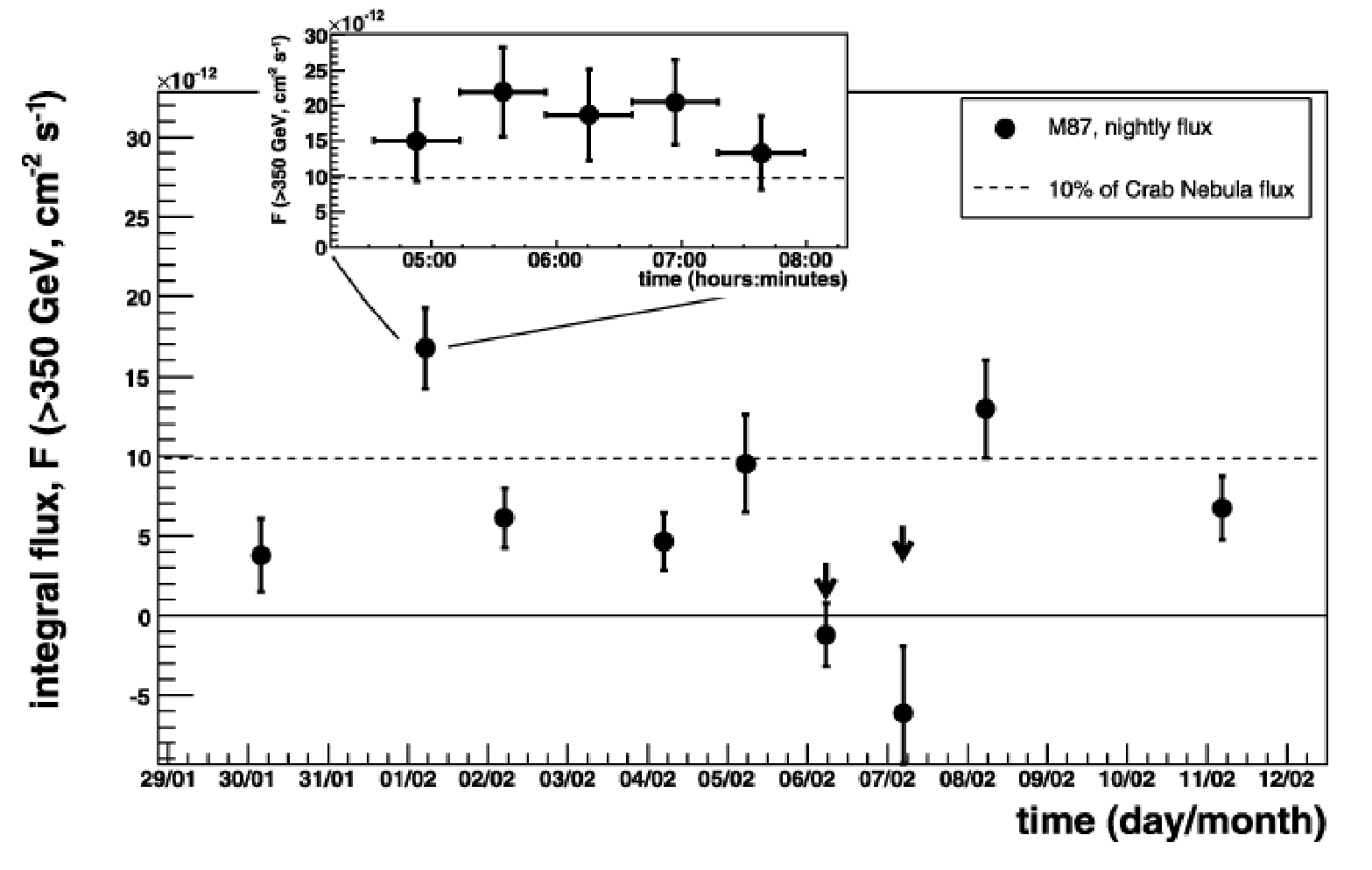}
  \caption{The night-by-night light curve for M\,87 as measured from 2008 January 30 (MJD 54495) to 2008 February 11 (MJD 54507). 
The upper panel shows the flux in the energy bin 150--350\,GeV, being consistent with a constant emission.
The lower panel shows the integral flux above 350\,GeV;
flux variations are apparent on variability timescales down to 1 day. 
The inlay of the lower panel shows the light curve above 350 GeV in a 40 min time binning for the night with the highest flux (2008 February 1).
The vertical arrows represent flux upper limits (95\% c.l.) for the nights with negative excesses.}
  \label{fig3}
 \end{figure}
 
Finally, the spectrum and the light curves were corrected for trigger inefficiencies due to higher discriminator thresholds during partial moon light and twilight conditions \cite{Albert2008c}. 
These corrections are on the order of 0\%--20\%. 
The data were also analyzed with an independent analysis yielding, within statistical errors, the same results.

\section{Results}
  \begin{figure}[!t]
  \centering
  \includegraphics[width=2.5in]{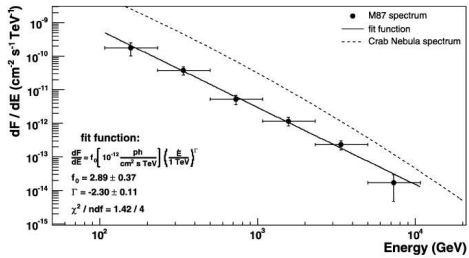}
  \caption{The differential energy spectrum of M\,87 for the total
data sample. The horizontal error bars represent width of the energy bins.
The best-fit function, assuming a power law, is given by the solid curve. 
The Crab Nebula spectrum is given by the dashed curve for reference.}
  \label{fig4}
 \end{figure}
 
  \begin{figure}[!t]
  \centering
  \includegraphics[width=2.5in]{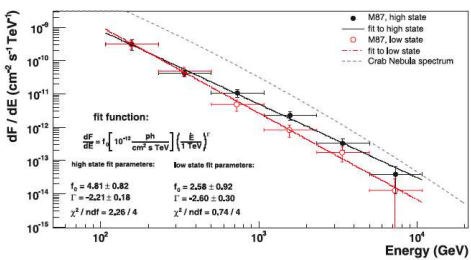}
  \caption{Differential energy spectra of M\,87 divided into
high (filled circles) and low (open circles) states. See text for the details.
The best-fit functions, assuming power laws, are
given by the black solid and red dashed-dotted curves,
respectively.}
  \label{fig5}
 \end{figure}
 
The $|ALPHA|$ distribution after so-called tight cuts is shown in Figure \ref{fig1}. 
The applied cuts are SIZE~$>$~450 photoelectrons and hadronness h~$<$~0.02. 
After the final $|ALPHA|$ cut (resulting in an overall cut efficiency of 37\%), the total signal of 241 events over 349 normalized background events corresponds to an excess with a significance of 9.9~$\sigma$ along equation
(17) in Li \& Ma \cite{Li}. 
The highest flux was observed on 2008 February 1 at a significance of 8.0~$\sigma$.
In searching for time variability, the data set was divided into nine subsets, one per observing night. 
In Figure \ref{fig2} we show both the light curve above the analysis threshold (150--350 GeV) and in the energy range at which MAGIC has the highest sensitivity for a variability search ($>$350 GeV). 
The low-energy range shows no significant variability with a 
$\chi^2_\mathrm{\nu}$ of 12.6/8 (probability of $P=0.13$)
for a constant fit. 
Instead, in the energy range above 350 GeV clear variability is found. 
A fit by a constant has a 
$\chi^2_\mathrm{\nu}$ of 47.8/8
corresponding to 
$P=1.1\cdot10^{-7}$. 
The correlation coefficient between the two energy bins is 
$r = -0.25^{+0.40}_{-0.33}$ (1-$\sigma$ errors), 
suggesting that there is no significant correlation, but we note rather large error bars in the low energy flux bin. 
We also investigated a night-by-night variability. 
There are five pairs of observations on consecutive nights in the total data set. 
We calculated individual probabilities $S_i$ for these pairs to have the same flux level and the corresponding significances.
We then computed a combined significance $S_{comb}$ \cite{Bityukov}: 
$ S_\mathrm{comb} = \left( {\sum{S_i}} \right) / {\sqrt{n}} $, with $n=5$. 
The resulting $S_\mathrm{comb} = 5.6\,\sigma$, which we interpret as a proof that the flux varies on timescales of 1 day or below.
Note that the 1 day variability is claimed from this combined analysis rather than from the 2008 February 1 flare alone.
We find our statistics not sufficient enough to determine the flare shape. 
Given the number of the observed changes in the flux level, the data belong to a complex of two, if not three, subflares.
We also looked for shorter time variability, but in none of the observation nights there is a significant flux variation in the two energy bands. 
A typical example in a 40 minute binning is shown in the inset in Figure \ref{fig2} for 2008 February 1.
The averaged differential energy spectrum of M\,87 (Fig. 3) extends from $\sim$100~GeV to $\sim$10~TeV and can be well approximated by a power law:
%\small
\[
{\frac{\mathrm{d}F}{\mathrm{d}E}}=(2.89\pm0.37)\times10^{-12}\left({\frac{E}{1\,\mathrm{TeV}}}\right)^{-2.30\pm0.11}\,
\frac 1 {\mbox{TeV}\,\mbox{cm}^{2}\,\mbox{s}}.
\]
%\normalsize
The errors are statistical only. 
We estimate an 11\% systematic uncertainty in the normalization and 0.20 for the spectral index \cite{Albert2008a}. 
The measured values are in good agreement with the H.E.S.S. (spectral index $\Gamma = -2.2 \pm 0.15$; Aharonian et al. 2006) and VERITAS ($\Gamma = -2.31 \pm 0.17$; Acciari et al. 2008) results. 
The observed spectrum is not significantly affected by the evolving extragalactic background light (EBL) due to the proximity of M\,87 \cite{NeronovAharonian}. 
To investigate a possible hardening of the spectrum with increasing absolute flux level, we divided the data sample into high and low state subsamples. 
The high sample comprises the two nights with the highest flux above 350 GeV (February 1 and 8), while the low state comprises the nights of lower flux data (January 30, February 2, 4, and 11). 
Both the high and low state spectra (Fig. 4) can be well described by a power law:
\[
{\frac{\mathrm{d}F}{\mathrm{d}E}}= f_0 \left({\frac{E}{1\,\mathrm{TeV}}}\right)^{\Gamma}\,
\left[ \frac {10^{-12}} {\mbox{cm}^{2}\,\mbox{s}\,\mbox{TeV}} \right]
\]
with $f_0^{\mathrm{high}} = (4.81 \pm 0.82)$, $\Gamma^{\mathrm{high}} = (-2.21 \pm 0.18)$
and   $f_0^{\mathrm{low}} = (2.58 \pm 0.92)$, $\Gamma^{\mathrm{low}} = (-2.60 \pm 0.30)$
for the \textit{high} and  \textit{low} states, respectively. 
There is a marginal hardening of the spectral index with the higher flux on the level of 1--2 standard deviations, depending on the way the significance is calculated. 
This hardening is not significant, which can be a consequence of the fact that the two flux levels (states) differ by less than a factor of~2.

\section{Discussion}
M\,87 is the first non-blazar radio galaxy known to emit VHE $\gamma$-rays and one of the best-studied extragalactic black hole systems. 
To enable long-term studies and assess the variability timescales of M\,87, the H.E.S.S., VERITAS, and MAGIC collaborations established a regular, shared monitoring of M\,87 and agreed on mutual alerts in case of a significant detection.
During the MAGIC observations, a strong signal of 8~$\sigma$ significance was found on 2008 February 1, triggering the other IACTs as well as Swift observations. 
For the first time, we assessed the energy spectrum below 250 GeV, where our observations can be well described by a power law that shows no hint of any flattening.
Our analysis revealed a variable (significance: 5.6~$\sigma$) night-to-night $\gamma$-ray flux above 350 GeV, while no variability was found in the 150--350 GeV range. 
We confirm the E~$>$~730 GeV short-time variability of M\,87 reported by Aharonian et al \cite{aha2}. 
The observed variability timescale is on the order of or even
below one day, restricting the emission region to a size of $R\leq
\Delta t\,c\,\delta = 2.6 \times 10^{15}\,\mathrm{cm} =
2.6\,\delta R_S$. 
The Doppler factor $\delta$ is only relevant for an emission region not expanding while traversing the jet. 
In case of an expanding-jet hypothesis, the {\em initial} radius of the expanding shell is given by $R^* < c\,\Delta t$ .
The emission can occur very close to the BH, provided that the ambient photon density is low enough as to allow the propagation of VHE $\gamma$-rays. 
Otherwise the emission region must be located farther away from the BH. 
In the latter case, the variability constraints can be met only if the emitting plasma does not substantially expand while traversing the jet, or if it moves with $\delta \geq 10$.
There exists no lower limit on the size of HST-1, and thus the flux variability cannot dismiss HST-1 as possible origin of the TeV flux. 
During the MAGIC observations, however, HST-1 was at a historically low X-ray flux level, whereas at the same time the luminosity of the M\,87 core was at a historical maximum (D. Harris 2008, private communication). 
This strongly supports the core of M\,87 as the VHE $\gamma$-ray emission region.
Our data alone cannot put strong constraints on VHE $\gamma$-ray emission models. 
The relatively hard VHE spectrum found for M\,87 ($\Gamma\approx -2.3$) is not unique among the extragalactic VHE $\gamma$-ray sources if one considers intrinsic spectra, i.e., EBL corrected. 
Also, we did not measure a high-energy spectral cutoff.
The found marginal spectral hardening may be interpreted as a similarity to other blazars detected at VHE, where such hardening has often been observed.
Our results show that a dense TeV monitoring, as exercised by ground-based IACTs, has revealed highly interesting rapid flares in M\,87. 
This fastest variability observed so far in TeV $\gamma$-rays observed in M\,87 restricts the size of the $\gamma$-emission region to the order of RS of the central BH of M\,87 and suggests the core of M\,87 rather than HST-1 as the origin of the TeV $\gamma$-rays. 
Results from the entire monitoring campaign, comprising data from other IACTs, will appear in a separate paper.

\section{Acknowledgement}
We would like to thank the Instituto de Astrofisica de 
Canarias for the excellent working conditions at the 
Observatorio del Roque de los Muchachos in La Palma. 
The support of the German BMBF and MPG, the Italian INFN 
and Spanish MCINN is gratefully acknowledged. 
This work was also supported by ETH Research Grant 
TH 34/043, by the Polish MNiSzW Grant N N203 390834, 
and by the YIP of the Helmholtz Gemeinschaft.

\end{document}